\documentclass{article}
\usepackage{spconf,amsmath,amssymb,graphicx,hyperref}
\hypersetup{
    colorlinks=true,
    urlcolor=blue,
    citecolor=black,
    linkcolor=black
}
\usepackage[OT1]{fontenc}
\usepackage{booktabs}
\usepackage{etoolbox}
\usepackage{balance}
\hypersetup{hidelinks}
\newcommand{\authoraffiliation}[1]{
  {\normalfont\fontsize{10}{12}\selectfont
  \begin{tabular}[t]{@{}c@{}}#1\end{tabular}}
}
\newcommand{\tablecaptiongap}{\par\vspace{5pt}}
\makeatletter

\patchcmd{\@maketitle}{\large \bf \@title}
  {\fontsize{14}{16}\selectfont\bfseries\@title}{}{}
\renewcommand{\footnotesize}{\@setfontsize\footnotesize{9}{11}}
\makeatother

\title{Structural Context from Audio and Lyric Evidence for Music Structure Analysis Using the New SheetSage-MSA Dataset}
\twoauthors
  {Hongbing Li}
  {\authoraffiliation{
   Queen Mary University of London, London, UK\\
   hongbingli96@gmail.com}}
  {Yaolong Ju$^*$\thanks{$^*$ Corresponding author.}}
  {\authoraffiliation{
   Great Bay University, Dongguan, China\\
   juyaolong@gbu.edu.cn}}
\begin{document}
\ninept
\maketitle
\begin{abstract}
Music structure analysis (MSA) aims to identify functional labels and their boundaries in music. Despite recent progress, three challenges remain.
First, fine-grained annotations for MSA are scarce. Second, existing systems often share one encoder between boundary detection and functional labeling despite their different information requirements. Third, most end-to-end systems are audio-only and ignore structural cues from lyrics. To address these issues, we construct SheetSage-MSA, an MSA dataset derived from SheetSage, containing 13,880 cleaned excerpt annotations from 9,010 songs and approximately 82.5 hours of annotated audio. We further propose SCALE (\textbf{S}tructural \textbf{C}ontext from \textbf{A}udio and \textbf{L}yric \textbf{E}vidence), which uses separate task-specific audio pathways to model local transitions and long-range structural context, and incorporates timestamped lyrics and lyrical recurrence to refine chorus predictions. On the HarmonixSet benchmark, SCALE outperforms the existing state-of-the-art SongFormer by 3.5 and 4.2 percentage points in HR.5F and HR3F, respectively, while improving ACC by 0.9 points. An interactive demo page containing the source code is also available at
\url{https://multimodal-music-research-lab.github.io/SCALE/}.

\end{abstract}
\section{Introduction}
\label{sec:intro}

\subsection{Background}
\label{ssec:subhead}

Music structure analysis (MSA) aims to divide a music recording into contiguous, non-overlapping sections and assign a label to each section \cite{nieto2020audio}. Two tasks are involved in MSA: \textbf{boundary detection} is to obtain the onset and offset for each section, and \textbf{functional labeling} annotates each section with a structural label. As MSA has developed, recent studies have gradually converged on several common choices. For function labels, many systems adopt the taxonomy of \cite{wang2022catch}, which consolidates heterogeneous annotations into seven structural functions: \texttt{intro}, \texttt{verse}, \texttt{chorus}, \texttt{bridge}, \texttt{outro}, \texttt{inst (instrumental)}, and \texttt{silence}. HarmonixSet \cite{nieto2019harmonix} has therefore been increasingly favored over earlier datasets such as SALAMI \cite{SALAMI} and RWC-Pop \cite{RWC}, as its labels align readily with this taxonomy. 

Model backbones have evolved in parallel. Earlier CNN-based systems used supervised learning on spectrogram inputs for boundary detection and structural feature learning  \cite{wang2021supervised,ullrich2014boundary}, while later work adopted attention-based architectures to better model long-range musical context \cite{wang2022catch,kim2023allinone,buisson2024linkseg}. More recently, pretrained music representations have become central to MSA~\cite{zhang2025temporal}. Rather than learning task-specific audio features solely from limited structural annotations, these systems transfer general musical representations learned from large-scale audio corpora~\cite{won2024musicfm,zhu2025muq}. SongFormer \cite{songformer} in particular further fuses short- and long-window self-supervised representations, allowing its Transformer to capture both fine-grained acoustic changes and long-range structural context.

Despite these advances, current MSA datasets and models still have clear limitations. First, fine-grained structural annotations remain scarce. HarmonixSet contains only 912 songs~\cite{nieto2019harmonix}, leaving limited supervised data after validation and test splits. Second, many recent systems share one encoder between boundary detection and functional labeling \cite{kim2023allinone,buisson2024linkseg,songformer}, although the two tasks require different information. Boundary detection depends strongly on precise timing and local musical changes, whereas functional labeling benefits more from longer musical context \cite{wang2022catch,kim2023allinone}. Third, most end-to-end systems remain audio-only and make limited use of non-audio structural cues. Lyrics, for example, are closely tied to song form: verses tend to
introduce new text, whereas choruses often repeat the same or similar
lines, making lyrical recurrence a useful structural cue. Although lyrics
have been used for chorus detection~\cite{bao2026almachor}, they have not
been integrated into end-to-end MSA for joint boundary and functional-label
prediction.

\subsection{Contributions}
\label{ssec:contributions}

\begin{figure}[t]
    \centering
    \includegraphics[width=0.95\columnwidth]
    {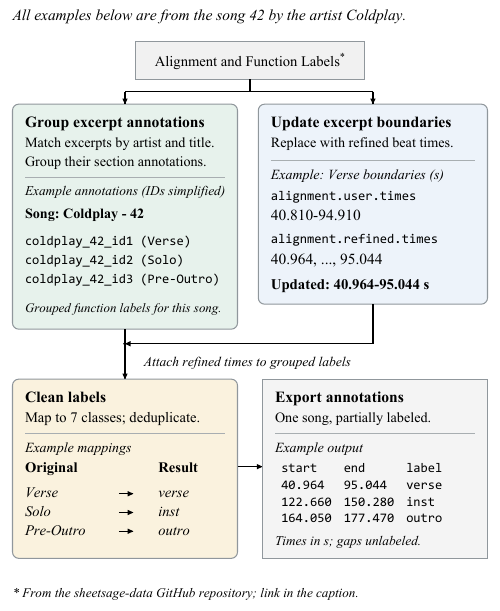}
    \caption{SheetSage-MSA data creation pipeline based on the
\href{https://github.com/chrisdonahue/sheetsage-data/}{\texttt{SheetSage-data}}
repository~\cite{donahue2022melody}. Audio is downloaded from YouTube using
links in
\href{https://github.com/chrisdonahue/sheetsage-data/blob/main/hooktheory/Hooktheory.json.gz}
{\texttt{Hooktheory.json.gz}}, which also provides alignments. Function labels
come from
\href{https://github.com/chrisdonahue/sheetsage-data/blob/main/hooktheory/Hooktheory_Raw.json.gz}
{\texttt{Hooktheory\_Raw.json.gz}}.}
    \label{fig:data_processing}
\end{figure}

In this work, we address the aforementioned three limitations. First, we present SheetSage-MSA, a new dataset for music structure analysis. For data acquisition, we follow the SheetSage-A2S
procedure~\cite{cummins2026sheetsagea2s} to obtain audio and metadata,
and then apply our own pipeline to derive MSA boundaries and labels. The resulting SheetSage-MSA dataset contains 13,880 annotated excerpts from 9,010 distinct songs, totaling approximately 82.5 hours. Second, we introduce SCALE, an MSA model with separate audio pathways for boundary detection and functional labeling. This design allows the two tasks to learn task-specific representations, while combining local transition cues with long-range structural context. Third, SCALE incorporates timestamped lyrics and uses lyrical recurrence to refine chorus prediction. On the SongFormBench-HarmonixSet (SongFormBench-HX) benchmark, SCALE establishes new state-of-the-art results across all three evaluation metrics, with gains over SongFormer~\cite{songformer} of 3.5 percentage points in HR.5F, 4.2 points in HR3F, and 0.9 points in ACC.

\section{Methods}
\label{sec:methods}

In this section, we first describe the construction of the SheetSage-MSA dataset, 
then present the SCALE architecture and training objective.

\subsection{Dataset}
\label{ssec:dataset}

We construct an MSA dataset from the function
labels and boundary annotations released with
SheetSage\footnote{\url{https://github.com/chrisdonahue/sheetsage-data/}}
~\cite{donahue2022melody}. Each entry describes an annotated excerpt
and includes its identifier, artist and title, user-specified timing,
and a beat-to-time alignment refined using
madmom~\cite{bock2016madmom}. Following the data-acquisition procedure
in~\cite{cummins2026sheetsagea2s}, we retrieve recordings from the
YouTube links in the metadata. Audio was available for 20,072 of the
26,175 excerpts; the remaining 6,103 could not be retrieved because
their source recordings were unavailable or copyright-restricted.
The subsequent processing is summarized in
Fig.~\ref{fig:data_processing}, containing the following steps.

\noindent\textbf{Group excerpt annotations.}
Multiple excerpts may correspond to different annotated regions of
the same song. We group all excerpts sharing the same artist and title
and aggregate their section annotations into a song-level annotation set.

\noindent\textbf{Update excerpt boundaries.}
For each excerpt, we
use the first and last beat timestamps in the refined alignment as its
start and end time, respectively. When beat timestamps are unavailable,
the user-specified end time is retained.

\noindent\textbf{Clean labels.}
We map the original function names to seven classes:
\texttt{intro}, \texttt{verse}, \texttt{pre-chorus}, \texttt{chorus},
\texttt{bridge},\allowbreak\ \texttt{instrumental}, and \texttt{outro}, 
as shown in
Fig.~\ref{fig:data_processing}. 
Annotations with missing,
composite, or unsupported function names are excluded, and duplicate
annotations are removed.

\noindent\textbf{Export annotations.}
The retained excerpt annotations are placed on the timeline of the
selected full-length recording. This produces partial song-level
structure annotations rather than complete annotations for every
recording. Regions without retained annotations remain unlabeled and
are masked out of the supervised losses.

The resulting dataset contains 13,880 cleaned excerpt-level section
annotations from 9,010 distinct songs, with a cumulative annotated
duration of approximately 82.5 hours. The dataset will be released under a CC BY-NC-SA 4.0 license.

\subsection{Model}
\label{ssec:model}

\begin{figure*}[t]
    \centering
    \includegraphics[width=\textwidth]{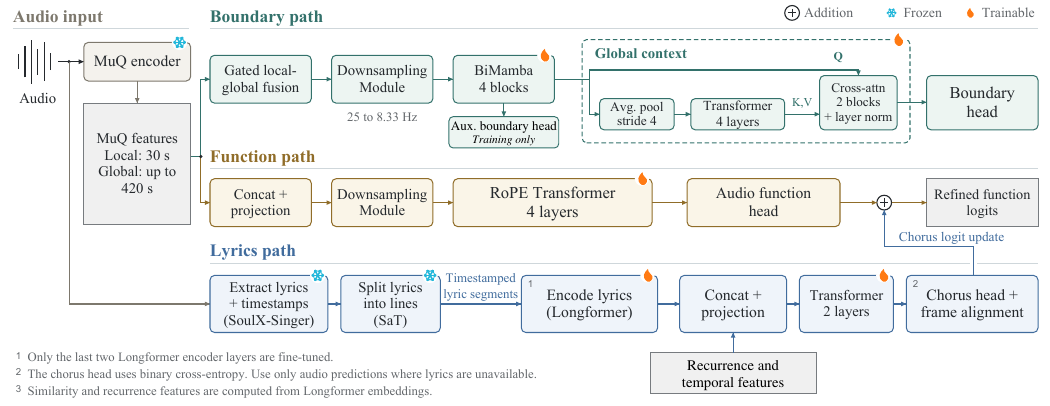}
    \caption{Overview of the proposed SCALE architecture. The model
comprises three paths: a boundary path, a function path, and a lyrics
path. MuQ~\cite{zhu2025muq} audio features are routed to the first two paths; the
boundary path combines BiMamba~\cite{gu2024mamba} with global Transformer~\cite{vaswani2017attention} context, while
the function path uses a RoPE~\cite{su2021roformer} Transformer for frame-level labels. The
lyrics path extracts and line-splits lyrics with SoulX-Singer~\cite{soulxsinger} and SaT~\cite{wtpsplit},
encodes them with Longformer~\cite{beltagy2020longformer}, and uses similarity and recurrence
features to refine the chorus logits.}
    \label{fig:model_architecture}
\end{figure*}

Fig.~\ref{fig:model_architecture} illustrates SCALE with three paths: boundary, function, and lyrics. The former two paths
use features from a frozen MuQ encoder~\cite{zhu2025muq}: local
representations from 30-s windows and global representations
from inputs of up to 420 seconds. The lyrics path refines chorus
prediction. Boundaries and function labels are combined into labeled
music sections during decoding.

\noindent\textbf{Boundary Path.}
Boundary detection requires both local acoustic changes and long-range
structural context~\cite{wang2022catch,kim2023allinone}. We project and
frame-align the local-window and full-song MuQ representations, then use
a learned gate to adaptively fuse the two streams. The gated sum is
concatenated with both aligned representations and projected into the
boundary feature space. Following the downsampling
design of SongFormer~\cite{songformer}, our modified module reduces the
sequence rate from 25~Hz to approximately 8.33~Hz. Four BiMamba blocks~\cite{gu2024mamba} model bidirectional temporal
dependencies, allowing the model to compare musical content before
and after a potential boundary. An auxiliary boundary head provides
additional supervision directly on the BiMamba output during training.
As shown in Fig.~\ref{fig:model_architecture}, the BiMamba output is
average-pooled with stride four and processed by a four-layer
Transformer~\cite{vaswani2017attention}. Two cross-attention blocks then update the frame-level features using
the global representation as keys and values. The resulting features
are passed to the boundary head.

\noindent\textbf{Function Path.}
The function path predicts function labels independently of boundary
detection. The temporally aligned MuQ representations are concatenated
and projected. After downsampling, a four-layer Transformer with rotary positional embeddings (RoPE)~\cite{su2021roformer} produces features for the audio function head, which outputs frame-level multiclass logits. RoPE encodes relative temporal
relationships between frames, supporting the modeling of section order
over long recordings.

\noindent\textbf{Lyrics Path.}
To obtain timestamped lyrics, we process each recording offline using
the SoulX-Singer preprocessing pipeline~\cite{soulxsinger}. It detects
vocals with their start and end times and transcribes the corresponding
lyrics. Each timestamped lyric passage forms a segment, which
SaT~\cite{wtpsplit} further divides into lines. For each segment, Longformer~\cite{beltagy2020longformer} jointly
encodes all lyric tokens with only its last
two layers fine-tuned. We pool the resulting token
representations over each line and over the full segment to obtain line
and segment embeddings, respectively. Choruses are often marked by repeated lyrics, so we derive recurrence
features, including similarity statistics, by comparing each segment
with non-adjacent segments. This avoids interpreting continuity between
neighboring lyrics as recurrence. We also compute temporal features
representing the segment's relative midpoint and duration within the
song. The recurrence and temporal features are concatenated with the
segment embedding and projected to the model dimension. A two-layer Transformer and a chorus
head then produce one chorus logit per segment.

The segment-level chorus logits are projected to audio frames according
to each segment's start and end times. Logits from overlapping segments
are averaged to obtain the frame-level lyric logit $s_t$, while $m_t$
indicates whether frame $t$ is covered by any lyric segment. We apply
a learned calibration, shared across all datasets, before fusion:
\begin{equation}
\Delta^{\mathrm{lyr}}_t
=
m_t \left(a s_t + b\right).
\end{equation}
As shown in Fig.~\ref{fig:model_architecture}, the calibrated residual
updates only the chorus logit:
\begin{equation}
z_{t,c}
=
z^{\mathrm{func}}_{t,c}
+
\mathbb{I}[c=\mathrm{chorus}]
\Delta^{\mathrm{lyr}}_t,
\end{equation}
where $z^{\mathrm{func}}_{t,c}$ and $z_{t,c}$ denote the Function Path
logit and the final logit, respectively, for class $c$ at frame $t$.
The residual is zero outside lyric coverage and does not directly
affect other class logits or boundary predictions.

\noindent\textbf{Training Objective and Decoding.}
We supervise both the main and auxiliary boundary heads using
Gaussian-smoothed targets and class-weighted binary cross-entropy.
Frame-level multiclass cross-entropy separately supervises the Function
Path logits and the Refined function logits. The chorus head is directly
supervised using segment-level weighted binary cross-entropy. A segment
receives a positive or negative target only when the proportion of
valid annotated chorus or non-chorus frames reaches the corresponding
threshold; otherwise, it is excluded. The Lyrics Path additionally
receives gradients from the loss on the Refined function logits through
the chorus-only residual. The complete objective is
\begin{equation}
\begin{aligned}
\mathcal{L}
={}&
\lambda_{\mathrm{ref}}\mathcal{L}_{\mathrm{ref}}
+\lambda_{\mathrm{FP}}\mathcal{L}_{\mathrm{FP}}\\
&+\lambda_{\mathrm{bnd}}
\left(
\mathcal{L}_{\mathrm{bnd}}
+\lambda_{\mathrm{aux}}\mathcal{L}_{\mathrm{aux}}
\right)
+\lambda_{\mathrm{chr}}\mathcal{L}_{\mathrm{chr}},
\end{aligned}
\end{equation}
where $\mathcal{L}_{\mathrm{ref}}$, $\mathcal{L}_{\mathrm{FP}}$,
$\mathcal{L}_{\mathrm{bnd}}$, $\mathcal{L}_{\mathrm{aux}}$, and
$\mathcal{L}_{\mathrm{chr}}$ denote the losses for the Refined function
logits, Function Path logits, boundary head, auxiliary boundary
head, and chorus head, respectively. The $\lambda$ terms weight the
corresponding objectives. Boundary and function losses are computed only on frames with the
corresponding annotations, allowing partially labeled samples to be
used during training. 

During inference, peak picking with a probability threshold and minimum
peak spacing is applied to the main boundary probabilities. The Refined
function logits are averaged within each predicted segment, and the
highest-scoring label allowed for the target dataset is selected.

\section{Experiments}
\label{sec:experiments}

\begin{table}[t]
    \centering
    \caption{Dataset splits used in this work. SongFormDB-HX is used for
training and validation, SheetSage-MSA for training
only, and SongFormBench-HX for testing. All values denote numbers of
songs.}
    \label{tab:dataset_statistics}
    \tablecaptiongap
    \small
    \setlength{\tabcolsep}{5pt}
    \begin{tabular}{l|ccc}
        \toprule
        \textbf{Dataset} &
        \textbf{Train} &
        \textbf{Val} &
        \textbf{Test} \\
        \midrule
        SongFormDB-HX~\cite{songformer}       & 512   & 200 & --  \\
        SheetSage-MSA (Proposed)    & 9,010 & --  & --  \\
        SongFormBench-HX~\cite{songformer}  & --    & --  & 200 \\
        \bottomrule
    \end{tabular}
\end{table}

\begin{table}[!t]
    \centering
    \caption{Performance on SongFormBench-HX. Upper: models retrained on the
same HX split. Lower: respective training settings, with baseline results
reported by SongFormer~\cite{songformer}.}
    \label{tab:main_results}
    \tablecaptiongap
    \small
    \setlength{\tabcolsep}{4.5pt}
    \renewcommand{\arraystretch}{1.05}
    \begin{tabular}{@{}lccc@{}}
        \toprule
        \textbf{Model} & \textbf{ACC} &
        \textbf{HR.5F} & \textbf{HR3F} \\
        \midrule
        \multicolumn{4}{@{}l}{\textit{Same HX training split}} \\
        \midrule
        All-In-One (HX)~\cite{kim2023allinone} & 0.704 & 0.530 & 0.704 \\
        LinkSeg (HX)~\cite{buisson2024linkseg} & 0.769 & 0.589 & 0.775 \\
        SongFormer (HX)~\cite{songformer} & 0.795 & 0.703 & 0.784 \\
        SCALE (HX) & \textbf{0.798} &
        \textbf{0.730} & \textbf{0.821} \\
        \midrule
        \multicolumn{4}{@{}l}{\textit{Respective training settings}} \\
        \midrule
        All-In-One~\cite{kim2023allinone} & 0.740 & 0.596 & 0.730 \\
        LinkSeg~\cite{buisson2024linkseg} & 0.780 & 0.630 & 0.762 \\
        SongFormer~\cite{songformer} & 0.807 & 0.696 & 0.780 \\
        SCALE (HX+SheetSage-MSA) & \textbf{0.816} & \textbf{0.731} & \textbf{0.822} \\
        \bottomrule
    \end{tabular}
\par\vspace{10pt}
\begingroup
    \centering
    \caption{Ablation results for SCALE on the SongFormBench-HX test set. SCALE denotes the complete
    model, and each subsequent row represents a variant with one SCALE
    component removed.}
    \label{tab:scale_ablation}
    \tablecaptiongap
    \small
    \setlength{\tabcolsep}{4.5pt}
    \renewcommand{\arraystretch}{1.05}
    \begin{tabular}{@{}lccc@{}}
        \toprule
        \textbf{Model variant} & \textbf{ACC} &
        \textbf{HR.5F} & \textbf{HR3F} \\
        \midrule
        SCALE & \textbf{0.816} & \textbf{0.731} & \textbf{0.822} \\
        \midrule
        w/o BiMamba & 0.807 & 0.671 & 0.753 \\
        w/o Function RoPE & 0.781 & 0.727 & 0.816 \\
        Shared Function Encoder & 0.805 & 0.706 & 0.790 \\
        w/o Lyrics & 0.808 & 0.731 & 0.820 \\
        Local Boundary Only & 0.807 & 0.730 & 0.822 \\
        \bottomrule
    \end{tabular}
\endgroup
\end{table}

\begin{sloppypar}
\noindent\textbf{Data Splits.}
Dataset splits used in this work are specified in Table~\ref{tab:dataset_statistics}: we use SongFormDB-HX\footnote{SongFormDB-HX was introduced with SongFormer~\cite{songformer} and is derived from HarmonixSet with rule-based annotation corrections. See \url{https://huggingface.co/datasets/ASLP-lab/SongFormDB}.} and SheetSage-MSA for training.
For SongFormDB-HX, we follow the training and validation splits used by
SongFormer~\cite{songformer}. Our proposed SheetSage-MSA dataset provides
additional training supervision for the full SCALE model. All evaluations
use the 200-song SongFormBench-HX test set.
\end{sloppypar}

\noindent\textbf{Evaluation Metrics.}
We report frame-wise functional-label accuracy (ACC) and boundary
F-measures with matching tolerances of 0.5 seconds (HR.5F) and 3 seconds
(HR3F)~\cite{songformer,raffel2014mireval}. Evaluation uses seven functional categories, with pre-chorus
mapped to verse~\cite{wang2022catch,songformer}. Scores are averaged over test songs.

\noindent\textbf{Feature and Model Configuration.}
MuQ features are extracted at 25~Hz and downsampled to approximately 8.33~Hz using a downsampling strategy inspired by SongFormer~\cite{songformer}. The boundary path operates on 256-dimensional representations,
while the function Transformer uses a hidden dimension of 512.
All Transformers use eight attention heads. MuQ is kept frozen, while
the last two layers of Longformer are fine-tuned. A dropout rate of
0.10 is used throughout the audio and lyric pathways.

\noindent\textbf{Optimization and Decoding.}
We use AdamW~\cite{loshchilov2019adamw} with a learning rate of $1.5\times10^{-4}$, a separate
Longformer learning rate of $10^{-5}$, and weight decay of $10^{-4}$.
A cosine learning-rate schedule is applied after 600 warm-up steps.
Gradient accumulation gives an effective batch size of eight, and
gradients are clipped to a norm of 1.0. During mixed-data training,
each HX training example is repeated four times to increase its
contribution. Boundary decoding uses a probability threshold of 0.35
and a minimum peak separation of 2 seconds.

\section{Results}
\label{sec:results}

\noindent\textbf{Model Performance.}
Table~\ref{tab:main_results} reports results under two experimental settings: all four models are (1) re-trained on the same HX split and (2) trained using their respective training data; results for the latter baselines are taken from the SongFormer paper~\cite{songformer}. In the former experiment, SCALE outperforms SongFormer by 2.7, 3.7, and 0.3 points in HR.5F,
HR3F, and ACC, respectively, demonstrating the effectiveness of the proposed model design under the same training data setting. In the latter experiment, adding SheetSage-MSA further improves
SCALE's ACC by 1.8 percentage points, demonstrating its value as a
complementary source of training data.

\noindent\textbf{Ablation Study.}
To further study the importance of relevant modules from the proposed SCALE architecture, Table~\ref{tab:scale_ablation} lists five model variants.
Among these variants, removing BiMamba yields the lowest boundary scores
(0.671 HR.5F and 0.753 HR3F), while removing function RoPE yields the lowest ACC
(0.781).
The \textit{Shared Function Encoder} variant feeds the boundary representation directly
to the function head, resulting in lower performance across all three metrics.
The \textit{Local Boundary Only} variant bypasses global context and feeds the BiMamba
output to the main boundary head, leading to a slight decrease in ACC to 0.807.

\begin{figure}[t]
    \centering
    \includegraphics[width=0.95\columnwidth]
    {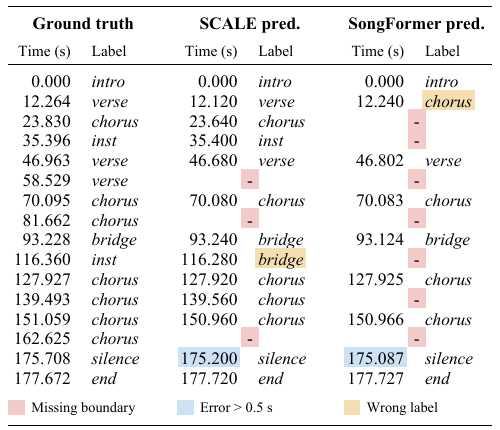}
    \caption{Analysis on \textit{Someday} by Tegan and Sara. SCALE recovers
more boundaries than SongFormer~\cite{songformer} but still misses some
boundaries between repeated sections with the same function label.}
    \label{fig:qualitative_analysis}
\end{figure}

\noindent\textbf{Qualitative Analysis.}
Figure~\ref{fig:qualitative_analysis} presents a representative comparison
on \textit{Someday} by Tegan and Sara. Consistent with its stronger
overall performance, SCALE recovers four boundaries missed by SongFormer
and avoids SongFormer's incorrect function label near 12.2~s. The dedicated Boundary Path helps localize these boundaries more
accurately, while evidence from repeated lyrics helps reject the
spurious chorus prediction. SCALE's main remaining limitation is boundary recall, particularly
between adjacent repetitions with the same function label, such as
consecutive verse or chorus sections. More fine-grained and consistent
annotations of repeated sections may improve recall.

\section{Conclusion and Future Work}
\label{sec:conclusion}

In this paper, we expand MSA training data with the proposed SheetSage-MSA dataset,
decouple boundary detection and functional labeling with task-specific
audio pathways, and incorporate lyrics as an additional
structural cue. These contributions enable SCALE to achieve state-of-the-art performance on SongFormBench-HX.

A remaining limitation is that existing datasets use different conventions for boundary annotation; for example, some boundaries are placed at vocal onset, while others are aligned
with the first beat of the measure. Such inconsistencies can affect both training and
evaluation. In future work, we plan to re-annotate a subset of the data
under a unified annotation protocol and use it as a standardized test set
to reduce evaluation bias caused by annotation conventions. We will also
explore scalable automatic methods to bring the training annotations
closer to this manually defined standard.

\section{Acknowledgment}
\label{sec:acknowledgment}
The computational resources are supported by SongShan Lake HPC Center (SSL-HPC) at Great Bay University.

\balance


\begin{thebibliography}{25}

\bibitem{nieto2020audio}
O.~Nieto, G.~J. Mysore, C.-i. Wang, J.~B.~L. Smith, J.~Schl\"uter,
T.~Grill, and B.~McFee, ``Audio-based music structure analysis: Current
trends, open challenges, and applications,'' \emph{Transactions of the
International Society for Music Information Retrieval}, vol.~3, no.~1,
pp.~246--263, 2020.

\bibitem{wang2022catch}
J.-C. Wang, Y.-N. Hung, and J.~B.~L. Smith, ``To catch a chorus, verse,
intro, or anything else: Analyzing a song with structural functions,''
in \emph{Proc. IEEE International Conference on Acoustics, Speech and
Signal Processing (ICASSP)}, 2022, pp.~416--420.

\bibitem{nieto2019harmonix}
O.~Nieto, M.~McCallum, M.~E.~P. Davies, A.~Robertson, A.~Stark, and
E.~Egozy, ``The Harmonix Set: Beats, downbeats, and functional segment
annotations of western popular music,'' in \emph{Proc. International
Society for Music Information Retrieval Conference (ISMIR)}, 2019,
pp.~565--572.

\bibitem{SALAMI}
J.~B.~L. Smith, J.~A. Burgoyne, I.~Fujinaga, D.~De~Roure, and
J.~S. Downie, ``Design and creation of a large-scale database of
structural annotations,'' in \emph{Proc. International Society for
Music Information Retrieval Conference (ISMIR)}, 2011, pp.~555--560.

\bibitem{RWC}
M.~Goto, H.~Hashiguchi, T.~Nishimura, and R.~Oka, ``RWC music database:
Popular, classical, and jazz music databases,'' in \emph{Proc.
International Conference on Music Information Retrieval (ISMIR)},
2002, pp.~287--288.

\bibitem{wang2021supervised}
J.-C. Wang, J.~B.~L. Smith, J.~Chen, X.~Song, and Y.~Wang,
``Supervised chorus detection for popular music using convolutional
neural network and multi-task learning,'' in \emph{Proc. IEEE
International Conference on Acoustics, Speech and Signal Processing
(ICASSP)}, 2021, pp.~566--570.

\bibitem{ullrich2014boundary}
K.~Ullrich, J.~Schl\"uter, and T.~Grill, ``Boundary detection in music
structure analysis using convolutional neural networks,'' in
\emph{Proc. International Society for Music Information Retrieval
Conference (ISMIR)}, 2014, pp.~417--422.

\bibitem{kim2023allinone}
T.~Kim and J.~Nam, ``All-in-one metrical and functional structure
analysis with neighborhood attentions on demixed audio,'' in
\emph{Proc. IEEE Workshop on Applications of Signal Processing to
Audio and Acoustics (WASPAA)}, 2023, pp.~1--5.

\bibitem{buisson2024linkseg}
M.~Buisson, B.~McFee, and S.~Essid, ``Using pairwise link prediction
and graph attention networks for music structure analysis,'' in
\emph{Proc. International Society for Music Information Retrieval
Conference (ISMIR)}, 2024, pp.~207--214.

\bibitem{zhang2025temporal}
Y.~Zhang, H.~Chen, J.-C. Wang, and J.~Chen, ``Temporal adaptation of
pre-trained foundation models for music structure analysis,'' in
\emph{Proc. IEEE Workshop on Applications of Signal Processing to
Audio and Acoustics (WASPAA)}, 2025.

\bibitem{won2024musicfm}
M.~Won, Y.-N. Hung, and D.~Le, ``A foundation model for music
informatics,'' in \emph{Proc. IEEE International Conference on
Acoustics, Speech and Signal Processing (ICASSP)}, 2024,
pp.~1226--1230.

\bibitem{zhu2025muq}
H.~Zhu, Y.~Zhou, H.~Chen, J.~Yu, Z.~Ma, R.~Gu, Y.~Luo, W.~Tan,
and X.~Chen, ``MuQ: Self-supervised music representation learning
with mel residual vector quantization,'' arXiv:2501.01108, 2025.

\bibitem{songformer}
C.~Hao, R.~Yuan, J.~Yao, Q.~Deng, X.~Bai, W.~Xue, and L.~Xie,
``SongFormer: Scaling music structure analysis with heterogeneous
supervision,'' arXiv:2510.02797, 2025.

\bibitem{bao2026almachor}
R.~Bao, H.~Ma, S.~Liu, C.~Gong, C.~Zhang, X.-L. Zhang, W.-Q. Zhang,
and X.~Li, ``ALMA-Chor: Leveraging audio-lyric alignment with Mamba
for chorus detection,'' in \emph{Proc. IEEE International Conference
on Acoustics, Speech and Signal Processing (ICASSP)}, 2026,
pp.~16102--16106.

\bibitem{donahue2022melody}
C.~Donahue, J.~Thickstun, and P.~Liang, ``Melody transcription via
generative pre-training,'' in \emph{Proc. International Society for
Music Information Retrieval Conference (ISMIR)}, 2022, pp.~485--492.

\bibitem{cummins2026sheetsagea2s}
E.~Cummins, Z.~Huang, A.~D'Hooge, Z.~Mo, and Y.~Ju, ``Audio-to-score
transcription using pre-trained features, data augmentation, and the
new SheetSage-A2S dataset,'' in \emph{Proc. ACM International Conference
on Multimedia (MM)}, 2026, to appear,
doi:~\href{https://doi.org/10.1145/3767308.3835653}{10.1145/3767308.3835653}.

\bibitem{bock2016madmom}
S.~B\"ock, F.~Korzeniowski, J.~Schl\"uter, F.~Krebs, and G.~Widmer,
``madmom: A new Python audio and music signal processing library,''
in \emph{Proc. ACM International Conference on Multimedia (MM)},
2016, pp.~1174--1178.

\bibitem{gu2024mamba}
A.~Gu and T.~Dao, ``Mamba: Linear-time sequence modeling with selective
state spaces,'' in \emph{Proc. Conference on Language Modeling
(COLM)}, 2024.

\bibitem{vaswani2017attention}
A.~Vaswani, N.~Shazeer, N.~Parmar, J.~Uszkoreit, L.~Jones,
A.~N. Gomez, L.~Kaiser, and I.~Polosukhin, ``Attention is all
you need,'' in \emph{Advances in Neural Information Processing
Systems}, vol.~30, 2017.

\bibitem{su2021roformer}
J.~Su, Y.~Lu, S.~Pan, A.~Murtadha, B.~Wen, and Y.~Liu,
``RoFormer: Enhanced Transformer with rotary position embedding,''
arXiv:2104.09864, 2021.

\bibitem{soulxsinger}
J.~Qian \emph{et~al.}, ``SoulX-Singer: Towards high-quality zero-shot
singing voice synthesis,'' arXiv:2602.07803, 2026.

\bibitem{wtpsplit}
M.~Frohmann, I.~Sterner, I.~Vuli\'c, B.~Minixhofer, and M.~Schedl,
``Segment any text: A universal approach for robust, efficient and
adaptable sentence segmentation,'' in \emph{Proc. Conference on
Empirical Methods in Natural Language Processing (EMNLP)}, 2024,
pp.~11908--11941.

\bibitem{beltagy2020longformer}
I.~Beltagy, M.~E. Peters, and A.~Cohan, ``Longformer: The long-document
Transformer,'' arXiv:2004.05150, 2020.

\bibitem{raffel2014mireval}
C.~Raffel, B.~McFee, E.~J. Humphrey, J.~Salamon, O.~Nieto,
D.~Liang, and D.~P.~W. Ellis, ``mir\_eval: A transparent implementation
of common MIR metrics,'' in \emph{Proc. International Society for
Music Information Retrieval Conference (ISMIR)}, 2014, pp.~367--372.

\bibitem{loshchilov2019adamw}
I.~Loshchilov and F.~Hutter, ``Decoupled weight decay regularization,''
in \emph{Proc. International Conference on Learning Representations
(ICLR)}, 2019.

\end{thebibliography}
\end{document}